# Label-free imaging of cholesterol and lipid distributions in model membranes

*Stephen H. Donaldson Jr.,\* Hilton B. de Aguiar\**

Département de Physique, Ecole Normale Supérieure / PSL Research University, CNRS, 24 rue Lhomond, 75005 Paris, France.

**Corresponding Authors**
\*steve.donaldson@phys.ens.fr; h.aguiar@phys.ens.fr.

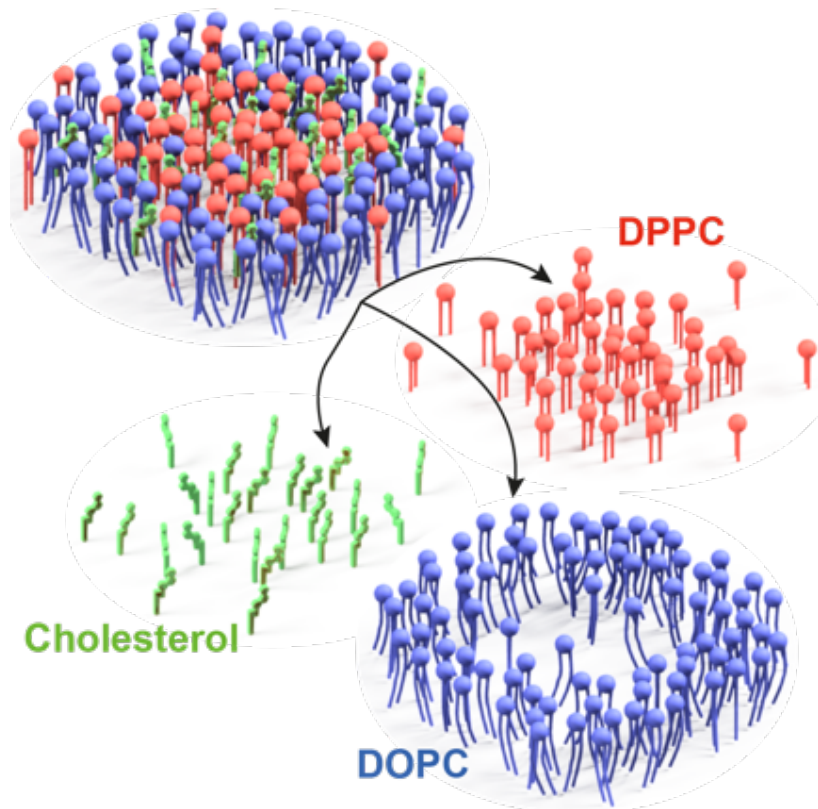

# ABSTRACT

Over recent decades, lipid membranes have become standard models for examining the biophysics and biochemistry of cell membranes. Interrogation of lipid domains within biomembranes is generally done with fluorescence microscopy *via* exogenous chemical probes. However, most fluorophores have limited partitioning tunability, with the majority segregating in the least biologically relevant domains (*i.e.,* low-density liquid domains). Therefore, a molecular-level picture of the majority of non-labeled lipids forming the membrane is still elusive. Here, we present simple, label-free imaging of domain formation in lipid monolayers, with chemical selectivity in unraveling lipid and cholesterol composition in all domain types. Exploiting conventional vibrational contrast in spontaneous Raman imaging, combined with chemometrics analysis, allows for examination of ternary systems containing saturated lipids, unsaturated lipids, and cholesterol. We confirm features commonly observed by fluorescence microscopy, and provide an unprecedented analysis of cholesterol distribution at the single-membrane level.

The cell membrane is the primary protective mechanical barrier between the cell interior and the outside world. Including a complex mixture of embedded proteins, lipids, cholesterol, receptors, and ion channels, it plays a crucial role in regulating molecular transport and signaling. Through lateral segregation of lipids, proteins, and cholesterol, liquid-liquid phase segregated lipid domains, also known as lipid rafts, are thought to participate in myriad membrane functions.[1–4] These roles include domain participation in key cell signaling processes such as immune responses or synaptic vesicle fusion,[5,6] or as pathological targets for viral or diseased protein binding at domains in HIV, Alzheimer's or cardiovascular disease.[2,3,7,8] Despite their apparent importance, direct detection of domains *in vivo* remains an intriguing challenge, largely due to lack of suitable instrumentation.[3]

Domains are simpler to observe and control *in vitro*, by using model supported membranes or vesicles. Indeed, studies of model membranes have revealed a rich set of biologically-relevant physicochemical behaviors for liquid-liquid domains,[9–12] as observed in reconstituted giant plasma membrane vesicles.[11] Liquid ordered (Lo) and liquid disordered (Ld) phases have been identified primarily by fluorescence microscopy, in which a fluorescent probe segregates into one of the liquid phases.[10,13] The Lo phase is generally thought to be enriched in cholesterol,[1–3,11] which is widely assumed to regulate the local compositions of both Lo and Ld phases.[3] However, quantitative *imaging* of cholesterol in domains remains challenging.

Considerable technical developments have been made for understanding lipid rafts. Fluorescence-based microscopies are the current workhorse for detecting lipid domains,[13,14] due to its sub-micrometer spatial resolution and fast speed. However, most fluorophores partition in the low-density domains,[3,15,16] and may lead to spurious phase diagrams.[17,18] Fluorescence microscopy strictly reflects the population of the fluorophores only, leaving open questions regarding the physico-chemical properties of high-density domains and their precise chemical composition. Nuclear magnetic resonance (NMR) has shown high chemical selectivity to map phase diagrams in multilamellar vesicles,[12] but has very poor spatial resolution. Other more recent label-free techniques, such as interferometric scattering (iSCAT)[19,20] and coherent Raman scattering (CRS)[21,22,23] microscopies, have shown much faster imaging capabilities of model membranes able to reach the acquisition rate necessary to detect the fast intermittent dynamics of lipid rafts in cell membranes. However, to date, there are no examples of spatially-resolved comprehensive chemical quantitation in relevant model ternary systems by CRS.[23–25] These are promising techniques for studying lipid rafts, but iSCAT does not have molecular selectivity – a crucial aspect to understand the specific interactions of cholesterol with lipids and proteins – and CRS microscopies are mostly based on cumbersome optical layout.

Spontaneous Raman microscopy allows chemical quantitation of lipids, and is therefore suitable for studying model membranes.[26] However, as the spontaneous Raman signal strength is typically weak (leading to low signal-to-noise ratios), it is difficult to perform chemically-resolved imaging containing various species with overlapping vibrational resonances in a single spectrum. Here, by exploiting tools from chemometrics,[27,28] together with suitable vibrational contrast,[29] we demonstrate sensitive chemical imaging of lipid monolayers using off-the-shelf chemicals in a single-shot basis (single spectrum), while avoiding chemical changes to the lipid structure for artificial signal enhancement.[15] These are remarkable achievements given that the imaging time used in our study is comparable to state-of-the-art CRS approaches applied to model membranes[21], however at generally lower interfacial density of lipids here (typical area per lipid > 60 Å$^2$).

Before imaging lipid mixtures, we first measured Raman spectra of pure lipid species. Fig. 1A presents the chemical structures of the three species used in this study, dipalmitoylphsphatidylcholine (DPPC), dioleoylphosphatidylcholine (DOPC), and cholesterol. As the three species have similar spectra in the C-H stretch region, in particular DOPC and DPPC, we used a deuterated version of DPPC (*d*-DPPC) to spectrally separate the ternary signals.[29] Fig. 1B shows the pure spectra of monolayers of the lipids and cholesterol. While the spectra of non-deuterated DPPC and DOPC are highly overlapping due to their nearly identical chemical structures,[15] the cholesterol spectrum presents more disparate features, in particular the lack of sharp resonances in the $CH_2$ stretch region (see Fig. 1B inset). Importantly, the spectrum of *d*-DPPC exhibits mostly C-D stretches, allowing *d*-DPPC to be spectrally distinct from DOPC and cholesterol. We then used these "pure" spectra as an input for a chemometrics-based analysis. We assume that the spectrum measured is a linear combination of basis spectra (pure monolayer spectra). As the number of spectral pixels is larger than the number of chemical species, it forms an overdetermined system of equations, allowing for straightforward inversion to retrieve their proportions to the measured spectrum (see methods for details).

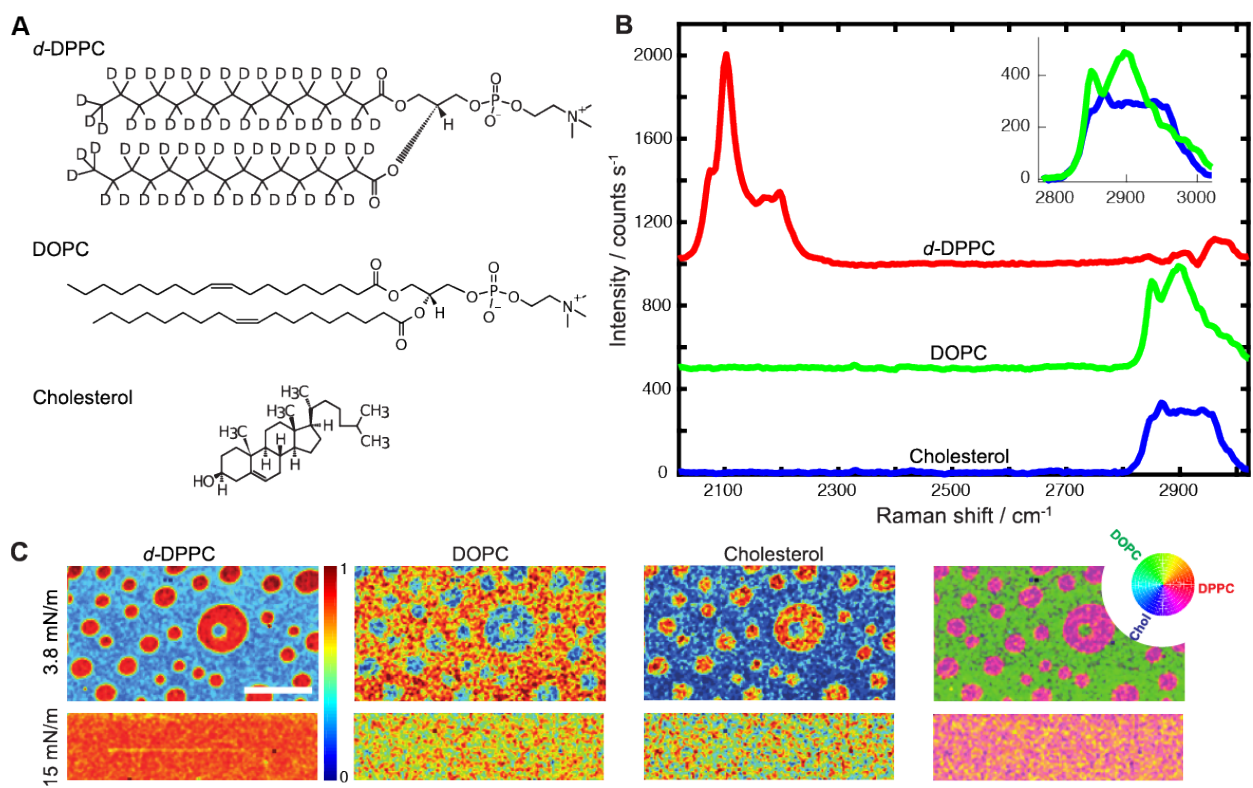

**Figure 1**. Single-shot chemical imaging of a ternary system. (A) Chemical structure of the species used in this study. (B) Raman spectrum of monolayers made of pure *d*-DPPC (upper red trace), DOPC (middle green trace), and Cholesterol (lower blue trace). The spectra have been offset for clarity reasons, however the inset compares DOPC and Cholesterol at the same scale. (C) Chemical images quantifying the local proportion of *d*-DPPC (first column panels), DOPC (second column panels), and Cholesterol (third column panels), in monolayer prepared at two pressures: 3.8 mN/m (upper panels) and 15 mN/m (lower panels). The merged false color image (fourth column panels) shows that the Lo domains are abundant in Cholesterol and DPPC, whereas the Ld phase is abundant in DOPC. The color wheel is a guide to the color mixing scheme. Scale bar: 10 μm.

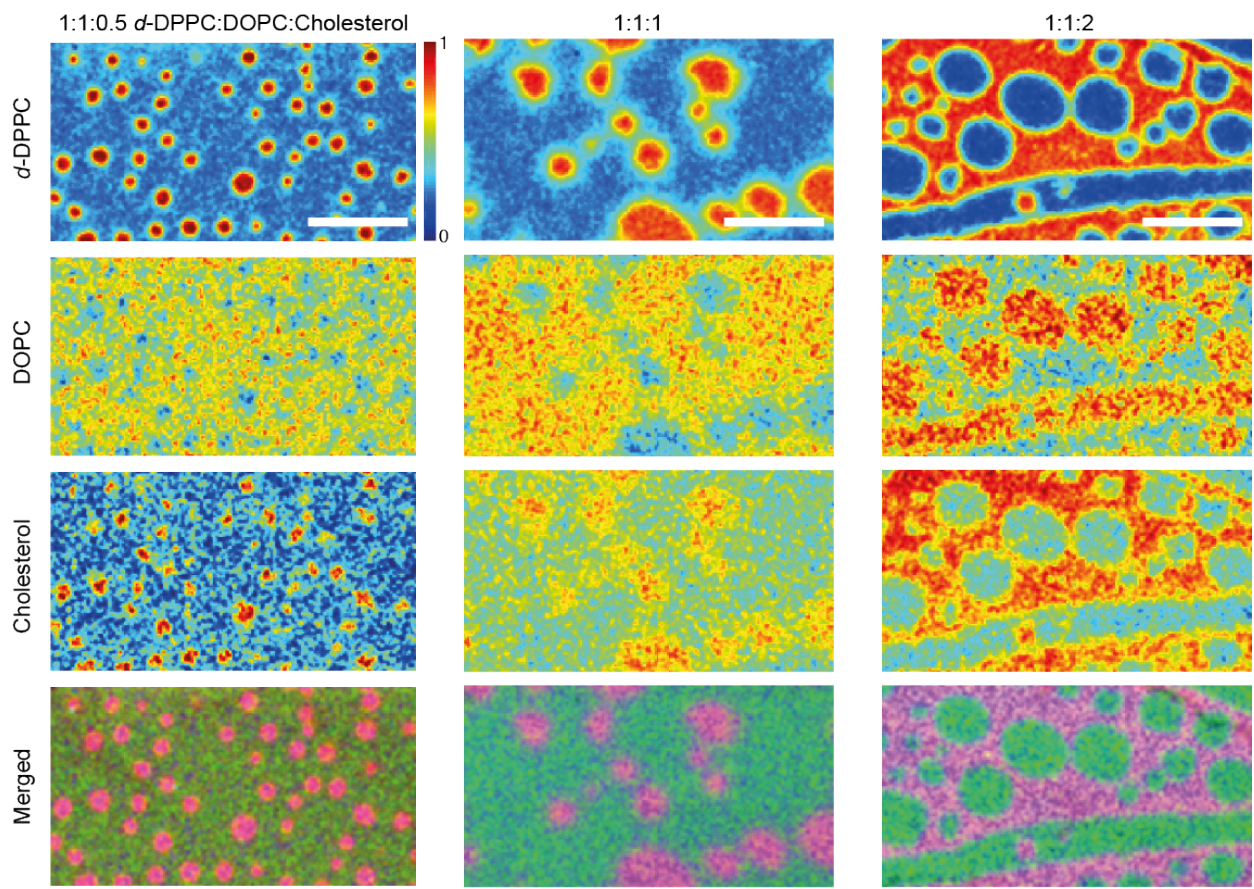

**Figure 2.** Chemical quantitation of ternary system upon adding cholesterol. The column images represent different cholesterol contents (left 20%, middle 33%, right 50%), and row images are for the various species quantified (DPPC, DOPC, Cholesterol, merged image in false colors according to Fig. 1C). Scale bars: 10 μm.

In the raft hypothesis,[3] Lo domains are believed to be formed of higher-density lipids embedded in a lower-density Ld phase. Therefore, we performed Raman-mode imaging on a classical ternary lipid mixture to chemically image the Lo and Ld phases. Figure 1C shows chemically-resolved images exploiting the chemometrics approach for a 1:1:1 *d*-DPPC:DOPC:Cholesterol system. At high surface pressure (~15 mN/m), no liquid-liquid phase separation is observed as the images are featureless up to the diffraction-limited resolution. However, at low surface pressure, the lipids phase separate into Lo and Ld phases. *d*-DPPC and cholesterol are enriched in circular Lo domains (bright regions in *d*-DPPC and cholesterol channels, Fig. 1C), while DOPC is depleted from the Lo domains (dark regions in DOPC channel, Fig. 1C) and is the dominant species in a sea of Ld phase (bright region in DOPC channel, Fig. 1C). Importantly, we note that every pixel corresponds to a single spectrum which provides spectral and spatial information on all three species in a single shot basis. The merged image (Fig. 1C, right-most panel) shows a straightforward qualitative inspection of the physical (topography) and chemical (relative partitioning) features combined together. The two pressures explored confirms previous phase diagrams for this system based on fluorescence imaging, in particular the existence of a phase transition between 3.8 mN/m and 15 mN/m. The phase transition was previously measured for this system at ~11 mN/m.[13]

Next, we altered the cholesterol content to observe composition-dependent morphology and local chemical composition of Lo-Ld phases. 3 compositions were tested, 1:1:0.5, 1:1:1, and 1:1:2 *d*-DPPC:DOPC:cholesterol, corresponding to ~20 mol%, ~33 mol% and ~50 mol% cholesterol. As shown in Figure 2, our results corroborate typical features of fluorescence-based studies, however with the additional capability of unambiguous chemical selectivity for both liquid phases. For the lowest cholesterol content (1:1:0.5), the morphology is similar to the 1:1:1 sample, however the domain sizes are smaller in the 1:1:0.5 case. Contrast inversion is observed for the 1:1:2 sample, *i.e.*, Ld domains enriched in DOPC are now dispersed in a sea of Lo, enriched in *d*-DPPC and cholesterol. The merged images (Fig. 2, bottom panels), clearly show that *d*-DPPC and cholesterol preferentially segregate together in the Lo phase throughout the phase space studied here.

These images allow for an approximate chemical quantitation of lipids and cholesterol domains. Based on the *d*-DPPC images, we rapidly acquired high signal-to-noise spectra at selected regions, either in the Lo or Ld phases identified from Fig. 2. Figure 3 presents the averaged spectra taken at the various positions of both domains for each composition. Overall, these spectra corroborate the analysis presented in the low signal-to-noise imaging presented in Fig. 1 and 2. In particular, (i) the "top-hat" spectral profile of cholesterol is dominant and continuously increases with the cholesterol concentration in the Lo domains; (ii) the predominance of DOPC in the Ld domains, evidenced by its sharper resonances (see the peaks at ~2850 cm$^{-1}$), and (iii) the correlation of the cholesterol spectral profile with the *d*-DPPC high-intensity one.

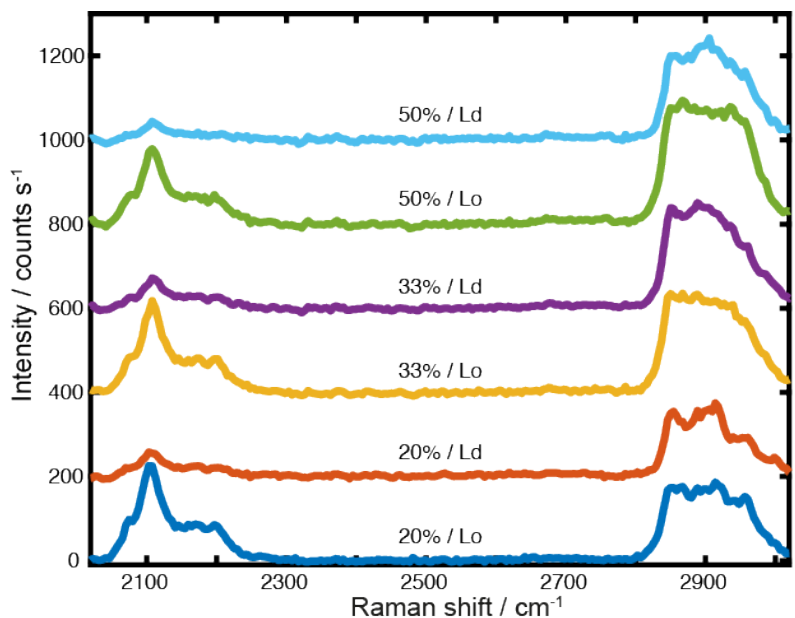

**Figure 3.** Low-noise spectra of lipid domains for various cholesterol content. Spectra have been offset for clarity.

To further probe the local interactions between cholesterol, *d*-DPPC, and DOPC, we quantify the local relative compositions of the three components within Lo and Ld phases. Table 1 presents an overview of the retrieved relative local proportions (or local surface density) for all samples, after the chemometrics analysis. In the Ld domains, the densities of *d*-DPPC and cholesterol are lower, whereas DOPC is higher, while the density of cholesterol increases in the Ld phase as the total cholesterol concentration increases. This observation may indicate that the Lo domains become saturated with cholesterol, and further addition of cholesterol partitions more into the Ld phase.

The relative local proportions from Table 1 can be used to calculate a free energy of transfer, $\Delta G$, between the lipid phases for each component,[30] as $\Delta G = -k_B T \ln K$, where $k_B$ is Boltzmann's constant, $T$ is temperature, and $K$ is the molar partition coefficient between the Lo/Ld phases reflected by the relative local composition shown in Table 1. Since the signal strength scales linearly with the local density, we directly use the relative composition ratio Lo/Ld for each species. $\Delta G$ values are also displayed in Table 1. A negative value indicates a preference for the Lo phase while a positive value indicates a preference for the Ld phase. $\Delta G$ of the $d$-DPPC is negative with magnitude generally greater than 1 $k_B T$ over the composition range studied here, indicating a strong preference for the Lo phase. On the other hand, $\Delta G$ of cholesterol is negative with large magnitude (> 1 $k_B T$) for low cholesterol composition, but the magnitude becomes less than 1 $k_B T$ for the higher cholesterol compositions, indicating that above a certain cholesterol content, the cholesterol no longer has a strong preference for the Lo phase. As for DOPC, $\Delta G$ is positive with magnitude generally less than 1 $k_B T$, indicating a weak preference for the Ld phase. These trends are in rough agreement with previous combined NMR and fluorescence measurements.[12,30] However, the local relative compositions and thermodynamics at the spatial resolution presented here cannot be straightforwardly accessed by other techniques.

**Table 1**. Relative local composition of lipids and cholesterol in the domains (Lo/Ld), and in parentheses the $\Delta G$ value in $k_B T$ units calculated as described in the text.

| Composition ($d$-DPPC:DOPC:cholesterol) | $d$-DPPC | DOPC | Cholesterol |
|---|---|---|---|
| 1:1:0.5 | 4.0±0.3 (-1.4) | 0.5±0.1 (0.7) | 3.9±0.8 (-1.4) |
| 1:1:1 | 3.1±0.3 (-1.1) | 0.6±0.1 (0.5) | 1.5±0.2 (-0.4) |
| 1:1:2 | 4.8±0.4 (-1.6) | 0.4±0.1 (0.9) | 2.2±0.1 (-0.8) |

In conclusion, we have accomplished chemically quantitative imaging of a ternary system with sub-micrometer resolution. This work paves the way for fast chemical analysis of lipid rafts using the emerging and faster Stimulated Raman Scattering (SRS),[31,32] because our methodology leads to straightforward spectral signatures for SRS microscopy.[28] In particular, the combination of compressive Raman[27,33] using optimized spectral filters masks with SRS,[34] could potentially lead to label-free chemical quantitation of cholesterol and lipids in realistic dynamic biological membrane rafts. To demonstrate the power of the developed imaging method, we made localized chemical and thermodynamic insights in both Lo and Ld phases over a range of biologically relevant cholesterol compositions. Upon increasing the cholesterol concentration, circular Lo domains grow in a sea of Ld until phase inversion occurs at ~40% cholesterol content, above which circular Ld domains grows in a sea of Lo. Our analysis confirms that Lo domains are enriched in $d$-DPPC and cholesterol, and further shows that DOPC is depleted from them, although not completely. With this chemically-selective imaging, all lipid phases (Lo, Ld, and solid phases) can be interrogated, and future directions include extending to lipid-protein interactions, lipid-lipid interactions, and membrane curvature effects.

**Experimental Methods**

**Supported lipid monolayer preparation**

DOPC and deuterated *d*-DPPC were purchased from Avanti, and Cholesterol from Sigma-Aldrich. Mixtures at the compositions described in the text were made in chloroform at a concentration of ~5 mg lipids/mL chloroform. All solution and sample preparation were done in a laminar flow hood. An adhesive imaging port was attached to a glass coverslip (#1.5 thickness). The coverslip was rinsed with ethanol, blown dry with $N_2$, and cleaned with high-RF ozone plasma for 7 minutes. After plasma cleaning, the coverslip was immediately immersed into a clean home-built Langmuir-Blodgett (LB) trough[35] filled with Milli-Q water. Several μL of the lipid solution were spread on the surface of the water and the solvent was allowed to evaporate for ~10 minutes. Then the barrier of the LB trough was controlled to the specified pressure, which was controlled through a feedback loop while the surface was pulled through the air-water interface to deposit a lipid monolayer. After about 5 minutes of drying in the laminar flow hood, a standard microscope slide (~1 mm thickness) was attached to the adhesive on the coverslip to seal the sample from the environment and to provide a rigid support during the imaging. Importantly, the lipid headgroups are strongly attached to the glass surface, allowing the domain morphology to be effectively frozen for imaging timescales. Domains become distorted several days after sample preparation, perhaps due to sample oxidation, so all samples shown in the text were freshly imaged immediately after the LB deposition.

**Raman microspectroscopy**

The microscope was a standard layout of an epi-detected Raman microscope. A pump laser beam (wavelength = 532 nm, Oxxius LCX-532) was spectrally cleaned-up by a bandpass-filter (FLH05532-4, Thorlabs), and its beam width expanded to 7.2 mm, before entering an inverted microscope (Nikon Eclipse Ti-U). Additional waveplates (half-waveplate and quarter-waveplate for 532 nm, Foctek Photonics) pre-compensated the ellipticity introduced by the dichroic filter (F38-532_T1, AHF), and also generated circularly polarized light to alleviate effects due to molecular orientation differences in the domains[36]. We used high numerical aperture (NA) oil-immersion objectives (Nikon 60x/1.4NA oil, figures 2 and 3, and Leica PL Fluotar 100x/1.30NA Oil PH3, figure 1) to ensure high-resolution imaging and increase collection efficiency. Pump power before the objective was 150 mW, power levels that ensured no degradation of domains within the scanned region. The samples were scanned with a high-resolution piezoelectric-based translation stage (Physik Instrumente P-545.3R7 stage, E-545.3RD controller). The Raman inelastic backscattered light was collected by the same objective and focused with the microscope tube lens on an anti-reflection-coated multimode fiber (M50L02S-A, Thorlabs, 50 um core), which also served as an effective pinhole for confocal detection. A notch filter at the fiber blocked residual pump light (NF533-17, Thorlabs), before guiding the signal to the spectrometer (Andor, Shamrock 303i, grating 300 l/mm), equipped with a high-sensitivity charge-coupled camera (Andor, iXon 897). All images presented were taken with integration times/pixel in the 0.1-0.3 s settings range.

## Data analysis

The spectral assignments for the lipids used in this study are well-established and can be found elsewhere[29,37]. All spectra were processed as raw data, using the following two-steps methodology. In the first step, we pre-processed each hyperspectrum using standard denoising routines of MATLAB (imgaussfilt3 function), because we have slightly oversampled in space and wavelength sampling domains. We have successfully tested that the 3D denoising procedure did not lead to artifacts in the image reconstructions, for instance, by using only a 2D denoising routine in each spatial map independently (imgaussfilt function). In the second step, we used concepts of chemometrics to obtain the final images: A spectrum $S$ can be written as a linear combination of eigenspectra (or pure components spectra) with a suitable weight. Mathematically, this can be written as $S = Pc$, where $P$ is a matrix with each column being the pure spectra of each species, and $c$ a vector representing the proportion of each component. Therefore, $c \cong (P^T P)^{-1} P^T S$, which is applied in a spatial-pixel-wise manner. Note that the pseudo inverse $(P^T P)^{-1} P^T$ is calculated only once and before retrieving all proportion images. For the analysis, we have considered as a fourth species the residual background Raman spectrum generated by the glass and the oil-immersion, together with the camera offset. The high signal-to-noise spectra were acquired by integrating for 5 s at selected spots within Lo and Ld after an overview imaging scan to identify the domain locations. The presented quantitative error values in Table 1 therefore reflect differences across the domains, which are relatively uniform in composition as observed in the images (Fig. 1C and 2).

## Acknowledgments

The authors thank Sylvain Gigan and Frederic Pincet for the equipment shared and fruitful discussions, Benneth Sturm for assistance with the data acquisition procedure, and Camille Scotte for discussions on the chemometrics analysis. The authors were supported by LabEX ENS-ICFP: ANR-10-LABX-0010/ANR-10-IDEX-0001-02 PSL*.

## References

(1) Lingwood, D.; Simons, K. Lipid Rafts As a Membrane-Organizing Principle. *Science* **2010**, *327* (5961), 46–50.
(2) Simons, K.; Ehehalt, R. Cholesterol, Lipid Rafts, and Disease. *J. Clin. Invest.* **2002**, *110* (5), 597–603.
(3) Sezgin, E.; Levental, I.; Mayor, S.; Eggeling, C. The Mystery of Membrane Organization: Composition, Regulation and Roles of Lipid Rafts. *Nat. Rev. Mol. Cell Biol.* **2017**, *18* (6), 361–374.
(4) Levental, I.; Veatch, S. L. The Continuing Mystery of Lipid Rafts. *J. Mol. Biol.* **2016**, *428* (24), 4749–4764.
(5) Varshney, P.; Yadav, V.; Saini, N. Lipid Rafts in Immune Signalling: Current Progress and Future Perspective. *Immunology* **2016**, *149* (1), 13–24.
(6) Milovanovic, D.; Honigmann, A.; Koike, S.; Göttfert, F.; Pähler, G.; Junius, M.; Müllar, S.; Diederichsen, U.; Janshoff, A.; Grubmüller, H.; et al. Hydrophobic Mismatch Sorts SNARE Proteins into Distinct Membrane Domains. *Nat. Commun.* **2015**, *6* (1), 5984.
(7) Yang, S.-T.; Kiessling, V.; Simmons, J. A.; White, J. M.; Tamm, L. K. HIV gp41–mediated Membrane Fusion Occurs at Edges of Cholesterol-Rich Lipid Domains. *Nat. Chem. Biol.* **2015**, *11* (6), 424–431.
(8) Mapstone, M.; Cheema, A. K.; Fiandaca, M. S.; Zhong, X.; Mhyre, T. R.; MacArthur, L.

H.; Hall, W. J.; Fisher, S. G.; Peterson, D. R.; Haley, J. M.; et al. Plasma Phospholipids Identify Antecedent Memory Impairment in Older Adults. *Nat. Med.* **2014**, *20* (4), 415–418.
(9) Dietrich, C.; Bagatolli, L. A.; Volovyk, Z. N.; Thompson, N. L.; Levi, M.; Jacobson, K.; Gratton, E. Lipid Rafts Reconstituted in Model Membranes. *Biophys. J.* **2001**, *80* (3), 1417–1428.
(10) Simons, K.; Vaz, W. L. C. Model Systems, Lipid Rafts, and Cell Membranes. *Annu. Rev. Biophys. Biomol. Struct.* **2004**, *33* (1), 269–295.
(11) Baumgart, T.; Hammond, A. T.; Sengupta, P.; Hess, S. T.; Holowka, D. A.; Baird, B. A.; Webb, W. W. Large-Scale Fluid/fluid Phase Separation of Proteins and Lipids in Giant Plasma Membrane Vesicles. *Proc. Natl. Acad. Sci. U. S. A.* **2007**, *104* (9), 3165–3170.
(12) Veatch, S. L.; Soubias, O.; Keller, S. L.; Gawrisch, K. Critical Fluctuations in Domain-Forming Lipid Mixtures. *Proc. Natl. Acad. Sci. U. S. A.* **2007**, *104* (45), 17650–17655.
(13) Veatch, S. L.; Keller, S. L. Organization in Lipid Membranes Containing Cholesterol. *Phys. Rev. Lett.* **2002**, *89* (26), 268101.
(14) Bagatolli, L. A.; Gratton, E. Two-Photon Fluorescence Microscopy Observation of Shape Changes at the Phase Transition in Phospholipid Giant Unilamellar Vesicles. *Biophys. J.* **1999**, *77* (4), 2090–2101.
(15) Ando, J.; Kinoshita, M.; Cui, J.; Yamakoshi, H.; Dodo, K.; Fujita, K.; Murata, M.; Sodeoka, M. Sphingomyelin Distribution in Lipid Rafts of Artificial Monolayer Membranes Visualized by Raman Microscopy. *Proc. Natl. Acad. Sci. U. S. A.* **2015**, *112* (15), 4558–4563.
(16) Shen, Y.; Zhao, Z.; Zhang, L.; Shi, L.; Shahriar, S.; Chan, R. B.; Di Paolo, G.; Min, W. Metabolic Activity Induces Membrane Phase Separation in Endoplasmic Reticulum. *Proc. Natl. Acad. Sci. U. S. A.* **2017**, *114* (51), 13394–13399.
(17) Veatch, S. L.; Leung, S. S. W.; Hancock, R. E. W.; Thewalt, J. L. Fluorescent Probes Alter Miscibility Phase Boundaries in Ternary Vesicles. *J. Phys. Chem. B* **2007**, *111* (3), 502–504.
(18) Skaug, M. J.; Longo, M. L.; Faller, R. The Impact of Texas Red on Lipid Bilayer Properties. *J. Phys. Chem. B* **2011**, *115* (26), 8500–8505.
(19) de Wit, G.; Danial, J. S. H.; Kukura, P.; Wallace, M. I. Dynamic Label-Free Imaging of Lipid Nanodomains. *Proc. Natl. Acad. Sci. U. S. A.* **2015**, *112* (40), 12299–12303.
(20) Wu, H. M.; Lin, Y. H.; Yen, T. C.; Hsieh, C. L. Nanoscopic Substructures of Raft-Mimetic Liquid-Ordered Membrane Domains Revealed by High-Speed Single-Particle Tracking. *Sci. Rep.* **2016**, *6* (1), 20542.
(21) Wurpel, G. W. H.; Schins, J. M.; Müller, M. Direct Measurement of Chain Order in Single Phospholipid Mono- and Bilayers with Multiplex CARS. *J. Phys. Chem. B* **2004**, *108* (11), 3400–3403.
(22) Potma, E. O.; Xie, X. S. Detection of Single Lipid Bilayers with Coherent Anti-Stokes Raman Scattering (CARS) Microscopy. *J. Raman Spectrosc.* **2003**, *34* (9), 642–650.
(23) Li, L.; Cheng, J. X. Label-Free Coherent Anti-Stokes Raman Scattering Imaging of Coexisting Lipid Domains in Single Bilayers. *J. Phys. Chem. B* **2008**, *112* (6), 1576–1579.
(24) Li, L.; Wang, H.; Cheng, J.-X. Quantitative Coherent Anti-Stokes Raman Scattering Imaging of Lipid Distribution in Coexisting Domains. *Biophys. J.* **2005**, *89* (5), 3480–3490.
(25) Potma, E. O.; Xie, X. S. Direct Visualization of Lipid Phase Segregation in Single Lipid Bilayers with Coherent Anti-Stokes Raman Scattering Microscopy. *ChemPhysChem* **2005**,

*6* (1), 77–79.
(26) Rieger, S.; Grill, D.; Gerke, V.; Fallnich, C. Quantitative Spontaneous Raman Scattering Spectroscopy in Artificial Binary Lipid Membranes. *J. Raman Spectrosc.* **2017**, *48* (10), 1264–1269.
(27) Wilcox, D. S.; Buzzard, G. T.; Lucier, B. J.; Rehrauer, O. G.; Wang, P.; Ben-Amotz, D. Digital Compressive Chemical Quantitation and Hyperspectral Imaging. *Analyst* **2013**, *138* (17), 4982–4990.
(28) Lu, F.-K.; Basu, S.; Igras, V.; Hoang, M. P.; Ji, M.; Fu, D.; Holtom, G. R.; Neel, V. A.; Freudiger, C. W.; Fisher, D. E.; et al. Label-Free DNA Imaging in Vivo with Stimulated Raman Scattering Microscopy. *Proc. Natl. Acad. Sci. U. S. A.* **2015**, *112* (37), 11624–11629.
(29) Gaber, B. P.; Yager, P.; Peticolas, W. L. Deuterated Phospholipids as Nonperturbing Components for Raman Studies of Biomembranes. *Biophys. J.* **1978**, *22* (2), 191.
(30) Veatch, S. L.; Polozov, I. V.; Gawrisch, K.; Keller, S. L. Liquid Domains in Vesicles Investigated by NMR and Fluorescence Microscopy. *Biophys. J.* **2004**, *86* (5), 2910–2922.
(31) Freudiger, C. W.; Min, W.; Saar, B. G.; Lu, S.; Holtom, G. R.; He, C.; Tsai, J. C.; Kang, J. X.; Xie, X. S. Label-Free Biomedical Imaging with High Sensitivity by Stimulated Raman Scattering Microscopy. *Science* **2008**, *322* (5909), 1857–1861.
(32) Nandakumar, P.; Kovalev, A.; Volkmer, A. Vibrational Imaging Based on Stimulated Raman Scattering Microscopy. *New J. Phys.* **2009**, *11*, 33026.
(33) Berto, P.; Scotté, C.; Galland, F.; Rigneault, H.; de Aguiar, H. B. Programmable Single-Pixel-Based Broadband Stimulated Raman Scattering. *Opt. Lett.* **2017**, *42* (9), 1696.
(34) Freudiger, C. W.; Min, W.; Holtom, G. R.; Xu, B.; Dantus, M.; Xie, X. S. Highly Specific Label-Free Molecular Imaging with Spectrally Tailored Excitation-Stimulated Raman Scattering (STE-SRS) Microscopy. *Nat. Photonics* **2011**, *5* (2), 103–109.
(35) Perez, E.; Wolfe, J. A Simple, Cheap, Clean, Reliable, Linear, Sensitive, Low-Drift Transducer for Surface Pressure. *Langmuir* **1994**, *10* (3), 974–975.
(36) Bélanger, E.; Bégin, S.; Laffray, S.; Koninck, Y. De; Vallée, R.; Côté, D. Quantitative Myelin Imaging with Coherent Anti-Stokes Raman Scattering Microscopy: Alleviating the Excitation Polarization Dependence with Circularly Polarized Laser Beams. *Opt. Express* **2009**, *17* (21), 18419–18432.
(37) Bunow, M. R.; Levin, I. W. Raman Spectra and Vibrational Assignments for Deuterated Membrane Lipids: 1, 2-Dipalmitoyl Phosphatidylcholine-d9 and-d62. *Biochim. Biophys. Acta (BBA)-Lipids Lipid Metab.* **1977**, *489* (2), 191–206.